\documentstyle[12pt,aasms4]{article}
\lefthead{Parravano \& S\'{a}nchez}
\righthead{Cloud Complexes Evolution}

\begin{document}

\title{Hierarchical Model for the Evolution of Cloud Complexes}

\author{N\'{e}stor M. S\'{a}nchez D.\altaffilmark{1,2} and
Antonio Parravano\altaffilmark{1}}
\affil{$^1$Centro de Astrof\'{\i}sica Te\'{o}rica, Facultad de Ciencias,\\
Universidad de Los Andes, M\'{e}rida, Venezuela.}
\affil{$^2$Departamento de F\'{\i}sica, Facultad Experimental de Ciencias,\\
Universidad del Zulia, Maracaibo, Venezuela.}

\begin{abstract}
The structure of cloud complexes appears to be well described by a
tree structure (i.e., a simplified ``stick man'') representation when
the image is partitioned into ``clouds''. In this representation, the
parent-child relationships are assigned according to containment.
Based on this picture, a hierarchical model for the evolution of
Cloud Complexes, 
including star formation, is constructed. 
The model follows the mass evolution of each
sub-structure by computing its mass exchange with its parent
and children. The parent-child mass exchange (evaporation or
condensation) depends on the radiation density at the interphase. 
At the end of the ``lineage'', stars may be born or die;
so that there is a non-stationary mass flow in the hierarchical 
structure.
For a variety of parameter sets the system follow the same series of steps 
to transform diffuse gas 
into stars, and the regulation of the mass flux in the tree by previously 
formed stars dominates the evolution of the star formation. 
For the set of parameters used here as a reference model, 
the system tends to produce 
IMFs that have a maximum at too high mass ($\sim 2 M_{\sun}$), and also 
the characteristic times for evolution seem too long. 
We show that these undesired properties can be improved by adjusting the model
parameters. The model requires further physics (e.g. allowing for 
multiple stellar systems and clump collisions) before a definitive comparison
with observations can be made. Instead, the emphasis here is to illustrate
some general properties of this kind of complex nonlinear model for the
star formation process.
Notwithstanding the simplifications involved, 
the model reveals an essential feature that will likely remain if
additional physical processes are included.
That is: 
the detailed behavior of the system is very sensitive to the
variations on the initial and external conditions, suggesting that
a ``universal'' IMF is very unlikely. 
When an ensemble of IMFs corresponding to a variety of initial or external 
conditions is examined, the slope of the IMF at high masses 
shows variations comparable to the range derived from observational data. 
These facts suggest that the considered physical processes 
(phase transitions regulated by the radiation field) may play a role 
in the global evolution of molecular complexes.

\end{abstract}


\section{Introduction}

There are both theoretical (Zuckerman \& Palmer 1974; Silk 1978;
Miyama et al. 1984) and observational
(Scalo 1985; Lada et al. 1991; Houlahan
\& Scalo 1992 and references therein) 
evidences that the formation of stars in molecular
complexes occurs by successive fragmentation in a hierarchical manner.
In fact, the maps of close complexes reveal clumps and irregularities 
on all observed scales and, inclusive, reveal some fractal
nature (Dickman et al. 1990; Falgarone et al. 1991), whose origin
is an important unsolved problem (see, e.g., the review of
Elmegreen 1993).
These complex structures are very difficult to model because of
the diversity of processes that are involved (i.e. MHD, complex 
chemical networks, radiative transfer in an inhomogeneous medium).
Many advances in the comprehension of these complex objects 
have been achieved by sectioning the problem and by 
focusing the attention on some scale and a few processes. 
However, in the present work we choose a schematic global approach, in
which the relationship between some processes occurring at different
scales can be examined.
When nonlinear processes are involved, these simple global models 
help to understand the response of the system to variations of 
initial and boundary conditions, and the sensitivity to variations 
of free parameters.

A way to schematize the structure of molecular complexes is
to consider that they are made of smaller sub-structures
frequently denominated as ``clouds''.
The definition of a ``cloud'' is ambiguous in the sense that it
depends on the molecular line
used for mapping and on the choice of the threshold intensity that
define the boundaries (Scalo 1990).
The hierarchical structure of these complexes can be described by
using the ``structure tree analysis'' introduced by Houlahan \&
Scalo (1992), which consists in a ``stick man'' representation
obtained by partitioning the image into clouds. In
the tree construction, a cloud is the ``child'' of another (of lower
density) if its boundary lies within its parent, and the
``skeleton image'' is build up by connecting parent to child clouds.
Based on a simplified version of this representation we elaborate 
a model to simulate
some aspects of the evolution of interstellar cloud complexes by 
computing the mass exchange
between parent and child clouds in a system of five levels of
hierarchy. At the end of the tree structure it is assumed that
stars may born
and die, affecting drastically the mass circulation in the ``tree''.

In \S 2 the model is described and the basic equations
are introduced. The results are discussed in \S 3 and,
finally, the main conclusions are summarized in \S 4.

\section{The model}

A mass exchange model in a hierarchical geometry is constructed
assuming a simplified Houlahan \& Scalo (1992) description:
that is: a) we kept only information on the lineage of sub-structures
(the Houlahan \& Scalo ``structure tree'' includes information on the
spatial positions as well, but we neglect spatial information in the 
present model),
and b) it is assumed that there are only five kinds of clouds
that could develop in the five predefined levels in the stick man
representation (the Houlahan \& Scalo representation is obtained
by successively thresholding the image at decreasing column
densities and identifying ``clouds'' as areas of connected pixels at
each grey level). 
These five levels represent, in
ascending order of density: warm gas ({\it wg}), neutral gas
({\it ng}), low density molecular gas ({\it lmg}), high density molecular 
gas ({\it hmg}) and protostars ({\it ps}).
The clouds at each level are assumed to
differ only by their mass and by the number and mass of
sub-structures they contain.
These idealizations obviously exclude many physical processes that
are known to occur (i.e., diffusion, collisions, etc.), but they allow
to perform the global study of this complex system.
We assume that each cloud can exchange mass with its parent cloud and
it may develop until ten interior sub-structures (hereafter children),
except at the last level 
where only one star may born from each protostar (that is, formation of
multiple star systems is excluded).
When a star dies, it incorporates a fraction of its mass to the
warm gas; the remaining fraction rests as a stellar remnant.
The first
level (warm gas) is subjected to an external radiation field $U_{o}$
and may accrete mass from the environment at a rate $A_{c}$.
Figure~1 schematizes the mass flux in a two children tree structure,
though a ten children structure is considered in this work (i.e. $10^{4}$
places to form stars).
The ideal number of available places for developing
children (the branching ratio)
should be large enough to ensure that every cloud always has a free place
to condense a child cloud if the conditions are given for it.
However, computational limitations restrict considering large branchings.
A branching ratio of $10$ is a compromise between computational time and
saturation. Even when saturation occurs in some substructures in
typical runs, a small fraction ($\sim 10\%$) of the total places in the
tree structure are occupied. On the other hand, Houlahan and Scalo (1992)
derived a value of $10$ for the average number of children for the
Taurus complex, although it is an uncertain result because it depends
on the chosen threshold intensities used to define the cloud boundaries 
and on the resolution of the maps.

Even if the mass exchange has a relatively low connectivity
(parent$\leftrightarrow$cloud$\leftrightarrow$ $10$ children),
it depends on the radiation transfer in the whole tree, which has a very
large connectivity.

Suppose that each cloud can exchange mass with its parent at a 
rate $\phi_{p}$, and with its {\it j}-th child
at a rate $\phi_{j}$,
then, the net
mass exchange rate of the {\it i}-th cloud with mass $M_{i}$ is
\begin{equation}
\frac{dM_{i}}{dt} = \phi_{p}(i) - \sum_{j} \phi_{j}(i)\ \ \ ,
\end{equation}
where $\phi>0$ ($\phi<0$) implies condensation (evaporation).
Mass conservation implies that if cloud $B$ is a child of cloud $A$,
then $\phi_{p}(B) = \phi_{j=B}(A)$.
Then, we
can schematize the mass exchange equations as: 
\begin{equation}
\begin{array}{lll}
M_{wg}(t+\Delta t) = M_{wg}(t) + \left(
                     A_{c} - \sum_{j} \phi_{j} \right) \Delta t
                     + \alpha \sum_{s} M_{s}
                     \delta (t-t_{s}-t_{*}(M_{s})) &
                     {\rm if\ level=1}\ \ \ ,\\
M_{i}(t+\Delta t)  = M_{i}(t) + \left(
                     \phi_{p} - \sum_{j} \phi_{j} \right) \Delta t &
                     {\rm if\ level=2,3,4}\ \ \ ,\\
M_{ps}(t+\Delta t) = M_{ps}(t) + \phi_{p} \Delta t -
                     M_{ps} \delta (M_{ps}-M_{cri}) &
                     {\rm if\ level=5}\ \ \ ;
\end{array}
\end{equation}
where $M_{s}$ represents the mass of the {\it s}-th star formed at the 
instant $t_{s}$ whose  lifetime is $t_{*}(M_{s})$, and $M_{cri}$,
the critical mass
required for a protostar of mass $M_{ps}$ to form a star, 
is a function of the history of the mass flux $\phi_{p}$
inward to the protostar (see below).
At the end of the life of a star it is assumed that
a mass $\alpha M_{s}$ is returned to the
warm gas phase, whereas the fraction $( 1-\alpha )M_{s}$ is left as a
stellar remnant. The dependence of $\alpha$ on the mass of the star
is assumed to be (David et al. 1990)
$\alpha= 1- 1.4/M_s$ if $M_s > 6 M_{\sun}$ and
$\alpha= 0.84- 0.44/M_s$ if $M_s < 6 M_{\sun}$.
However, this parameter has little influence on the behavior of the model
because the mass in stars is a small fraction of the mass of the system.
 
The mass exchange rate $\phi$ plays a central role in the model.
Many physical processes mediate to determine its value which depends
mainly on the unbalance between cooling and heating rates as well as
on the unbalance between formation and destruction rates of chemical species.
These rates depend on the local state of the gas
and on the external conditions. In particular, the far UV radiation
received by the cloud is an important factor in determining the heating rate
and the dissociating rate of molecular species.
For a given chemical state,
the thermal evolution of a gas 
depends mainly on the excess of heating or cooling.
If the pressure remains constant and
cooling dominates over heating, the gas temperature decreases and
the density increases.
However, during the thermal evolution, both the chemical state of the gas
and the cooling and heating efficiencies vary; generally a denser and
cooler thermo-chemical equilibrium is attained: that is, a phase transition
is achieved.
The dominating physical processes that controls the mass exchange
rate $\phi$ are different for the various transitions between the phases
considered in the model.
As an example, for the transition between the warm and neutral phases, the
relevant physical processes are the cooling by
the collisional excitation of heavy ions, the heating by photoejection of
electrons by dust, and the ionization of hydrogen by collisions and
by cosmic rays.
On the other hand, a relevant process for the transition between the
neutral and the low density molecular phases is the photodissociation by
UV radiation of molecular species such as ${\rm H_2}$ and ${\rm CO}$.

The present model only takes care of two variables:
the mass $M_i$ of substructures and the radiation density $U_i$
at their surfaces. Therefore we are restricted to adopt a
schematic dependence of $\phi$ that involves only the location
of the cloud in the tree structure and these two variables.
Within this restriction, we assume that the mass exchange rate
$\phi_{p}(M_{i},U_{i})$ between the substructure $i$ (located at level $l$
and being the $j$~-th brother) and its parent cloud (located at level $l-1$)
can be approximated by the product
\begin{equation}
\phi_{p}(M_{i},U_{i}) = \phi_{l} \; F(U_{i},l,j) \; G(M_{i},l)
\ \ \ .
\end{equation}

The first term $\phi_{l}$ is assumed to depend only
on the cloud level $l$
and represents the typical rate of condensation of a cloud of ``type
$l$'' contained in a parent cloud of ``type $l-1$'' when it has a typical
mass $M_{o,l}$ and no radiation field is present
(i.e. $\phi_{p}(M_{o,l},0)=\phi_{l}$, $F(0,l,j)=1$, and $G(M_{o,l},l)=1$).
The adopted values of $\phi_{l}$ in Table~1 correspond to
characteristic times $t_l=M_{o,l}/\phi_{l}$ of 0.67, 0.2, 0.02, and 0.2
Myr for $l =$ 2, 3, 4, and 5 respectively.
It is to be noticed that the time to condense a cloud of mass
$M_{o,l}$ is much longer that $t_l$ because $G(M_{i},l)<<1$ when
$M_{i}<<M_{o,l}$ (see below).

The second term takes into account the dependence of $\phi$ on the
radiation density $U_{i}$ at the surface of a cloud located in the
position $(l,j)$ in the tree structure.
In eq.~3 $F(U_{i},l,j)$ is the only factor that may change sign.
Function $F$ is assumed to be positive for low values of $U_i$
(i.e. mass transfer to cloud $i$ from its parent cloud)
and negative for high values of $U_i$ (i.e. mass transfer from cloud $i$ to
its parent cloud).
The adopted form for $F$ is
\begin{equation}
F(U_{i},l,j) = \left( 1 - \frac{U_{i}}{U_{cri}(l,j)} \right) \ \ \ ,
\end{equation}
where,
\begin{equation}
U_{cri}(l,j) = U_{l} / j^{a_{l}} \ \ \ .
\end{equation}
That is,  $F$ is a linear function of $U_i$ which
acts as a switch to allow evaporation or condensation depending
on the ratio of the radiation density $U_{i}$ to
a critical value $U_{cri}(l,j)$. This critical value depends mainly
on the level $l$ at which the cloud $i$ is located and, in a lesser degree,
on the position occupied by cloud $i$ in relation to their brothers clouds
(i.e. its $j$ number).

In practice, the direction of the phase transition
between a diffuse state (the parent cloud) and
a denser one (the child cloud)
not only depends on the radiation density
but also on the local state, on the composition of the gas, and on others
sources of heating, ionization or dissociation.
However, their combined effect can be included in $U_{cri}(l,j)$
if the physical conditions in clouds of the same type $(l,j)$
are assumed to be identical.
The values for $U_{l}$ in eq.~5 can be estimated from the typical values
of the critical energy density for the transition between the
``ISM phases''.
For example, for the typical ISM pressures and gas composition
in the solar vicinity, the critical energy density $U_2$
(corresponding to the transition from the {\it wg} to the first {\it ng}
child; $l=2$ and
$j=1$), has been estimated to be
$\sim 7 \times 10^{-17} erg \, cm^{-3} \, \AA ^{-1}$ in the band
between $912 - 1100  \, \AA$ (Spitzer 1978; Parravano 1988).
This is the relevant band at which $U_{cri}$ must be calculated
because this radiation is the main source of
heating in these two phases. Also, this radiation is the main responsible
of the photodissociation and heating in the molecular phases.
For the purposes of this model, all the values for $U_l$ can be
normalized to the value $U_2=1$.
To estimate $U_l$ for the other transitions, specific
models of photodissociation regions can be used (Hollenbach 1990,
and references therein).
If $\tau _l$ is the typical optical depth for the far UV radiation measured
from the {\it wg}-{\it ng} interphase to the interphase of the
phase $l$, then $U_l$ can be estimated as $U_l \sim U_2 \; \exp(-\tau _l)$.
The values of $U_l$ in Table~1 correspond to $\tau _3=0.7$,
$\tau _4=8.5$ and $\tau _5=9.2$.
Then, to transform the
warm gas into protostars, the clouds at each level must grow
until their optical depth is large enough to attenuate the radiation below
the corresponding critical value to permit the development of child clouds.
The weak dependence of $U_{cri}$ on $j$ (the child number) is introduced in
order to account for spatial variations on the physical conditions
inside parent clouds, such as the pressure and the opacity.
The adopted values of $a_l$ in Table~1
correspond to a difference in $U_{cri}$ between the first and
the 10-th brother of 2\% for levels 2, 3, and 4, and a difference
of 17\% for level 5. The chosen values of these small differences in
$U_{cri}$ for the brothers
clouds are arbitrary, but they are necessary to avoid
the simultaneous formation
and the identical evolution of all the child clouds in a given parent cloud.

The third term in eq.~3 takes into account the dependence of $\phi$ on the
area of the interphase between the cloud and its parent.
The adopted form for $G$ is
\begin{equation}
G(M_{i},l) = \left( \frac{M_{i}}{M_{o,l}} +
\delta \right) ^{2/3} \ \ \ .
\end{equation}
However, function $G$ is set to zero when the condensation is switch-on
(i.e. $F>0$) but the parent cloud is empty.
The $2/3$ exponent comes from the assumption that $\phi$ is
proportional to the surface of a spherical cloud of uniform density.
The term $\delta=10^{-3}$ is introduced in order to initiate
cloud condensation when the parent has available mass.
Notice from eqs.~3 and 6 that if the ratio $\phi_l/(M_{o,l}) ^{2/3}$ 
is kept constant, the parameter $M_{o,l}$ has little effect on $\phi$.

The regulating term $F(U_{i},l,j)$
couples the local mass flux with the stars and gas distributions 
in the whole system via the radiation transfer in the tree.
Thus, the radiation field controls the mass exchange in the system 
and regulates the star formation process.
There are three contributions
($U_{in}(i) + U_{out}(i) + U_{ext}(i)$)
to the radiation density ($U_{i}$) in the {\it i}-th 
cloud:\\
- First, the radiation density $U_{in}$ corresponding
to the radiation coming from the 
stars living in the substructures that are contained by the cloud.
It is calculated recursively beginning from the fourth level.
Due to the fact that there is not a protostar containing any star, 
the fifth level is not taken into account, and the interphase radiation 
density for a cloud at the fourth level is assumed to be
\begin{equation}
U_{in}(i) = \frac{e^{-\tau_{i}}}{D_{l}} \sum_{s} R_{s}\ \ \ ,
\end{equation}
where $\tau_{i}$ is the optical depth of the cloud, $D_{l}$
represents the dilution factor, and 
$R_{s}=r_o M_{s}^{3.5}$ 
represents the radiation density associated to the star $s$ inside the 
{\it i}-th {\it hmg} cloud.
The optical depth is assumed to be
$\tau_{i} = c_{o} n_{l}^{2/3} M_{i}^{1/3}$
(i.e, proportional to the cloud size), 
$n_{l}$ being the number density at the level $l$ (see Table~1).
The parameters $r_o$
and $c_{o}$ are fixed respectively to $6.3\times 10^{-3}$ and 
$1.3\times 10^{-2}$ in the Reference Model (see below).
The dilution factor is also assumed to depend only on the level as
the ratio of the typical cloud surfaces at consecutive levels 
(i.e. $D_{l}=[(n_{l+1} M_{o,l})/(n_{l} M_{o,l+1})]^{2/3}$.

For the superior levels ($l=3,2,1$), $U_{in}(i)$ is calculated by summing
the radiation $U_{in}(j)$ of their children attenuated by the optical
depth $\tau_{i}$ of the cloud and diluted by the factor $D_{l}$. This is,
\begin{equation}
U_{in}(i) = \frac{e^{-\tau_{i}}}{D_{l}} \sum_{j} U_{in}(j)\ \ \ .
\end{equation}
\\
- Second, the radiation density $U_{out}$ corresponding to the radiation 
coming from the stars in the system that are not contained by the cloud.
Obviously, $U_{out}(wg) = 0$. For the clouds at the levels 2, 3, 4
and 5, $U_{out}$ can be calculated recursively
beginning with the second level. That is,
\begin{equation}
U_{out}(i) = \left[ U_{out}(p) + \sum_{k} U_{in}(k) \right]
             \exp (-\tau_{p})\ \ \ ,
\end{equation}
where indices $p$ and $k$ indicate that the variables are evaluated on the 
parent or on the brothers clouds, respectively.\\
- Third, the
attenuated external radiation $U_{ext}(i) = U_{o}\exp (-\tau_{oi})$
that corresponds to the external radiation $U_{o}$ attenuated by the total
optical depth $\tau_{oi}$ from the surface of {\it i}-th cloud to the
surface of the system.
In order to explore the dependence of the system evolution
on the external radiation field $U_{o}$, it is useful to introduce
the quantity $\tilde{U} = U_{o}\exp (-\tau_{o,wg}) / U_{cri}(2,1)$, where
$\tau_{o,wg}$ is the initial optical depth of the warm gas cloud and
$U_{cri}(2,1)=U_{2}=1$ is the surface critical radiation for the first warm gas 
child. If $\tilde{U} > 1$, there is no condensation of warm gas even 
to its first child, and therefore, an initially warm gas remains as 
it is forever. The cases where $\tilde{U}$ is below but close to unity 
are of particular interest because it is expected that real ``{\it wg} 
clouds'' start to develop internal structures when the external 
condition changes from $\tilde{U} > 1$ to $\tilde{U} < 1$.
This change may be produced, for example, by a decrease of the external
radiation field or an increase of the warm gas optical depth due to
an increase of the system mass or the external pressure. However,
abrupt changes, due for example to cloud collisions or
shock fronts, may reduce $\tilde{U}$ drastically below unity.

The star formation process is very difficult to model because it involves
many interrelated physical processes. Moreover,
the basic mechanism that determines the mass of a star has
not yet been clearly established (see, for example, the reviews
of Shu et al. 1987 and Pudritz et al. 1997).
In the present model, we suppose that stars determine their own masses 
through the action of stellar winds (Adams \& Fatuzzo 1996).
The mass of the formed star can be calculated as a function of the 
formation time ($t_{f}$), i.e., the
time interval that the protostar has been accreting mass from the
parent cloud until the stellar wind stops the infall.
The dashed line in Figure~2 corresponds to the mass of the star formed
versus the formation time when the Adams \& Fatuzzo's (1996)
conditions for stopping the accretion are used, except that it is
assumed that the smallest possible stellar mass is $0.1 M_{\sun}$. 
The smooth curve shows the simple functional form assumed in this 
model, i.e.,
\begin{equation}
\label{mcri}
M_{cri} = \frac{C}{t_{f}^{\beta}} + M_{min}\ \ \ ,
\end{equation}
where $C = 100$, $\beta = 11/4$ and $M_{min} = 0.1 M_{\sun}$. The
dotted line in Figure~2 represents an example
of the mass evolution of a protostar that is
accreting mass at a variable rate from its parent cloud.
When $M_{ps}=M_{cri}$, the wind from the protostar is assumed to stop 
the accretion and a star of mass $M_{s}=M_{ps}$ appears at the
place of the protostar.
Therefore, as the accretion rate increases, the mass of the formed 
star increases and the formation time decreases.

Finally, each star is assumed to live a time $t_{*}$ 
given by (David et al. 1990)
\begin{equation}
\log{t_{*}(M_{s})} = 4.0 - 3.42\log{M_{s}} + 0.88(\log{M_{s}})^{2}\ \ \ ,
\end{equation}
where $M_{s}$ is in solar masses and $t_{*}$ in Myr. When a star
dies, the localization of that star remains free to form another star.

\section{Results}

Only some aspects of the evolutionary process that are expected to be
present in a molecular complex have been schematically considered in the 
above model.
Therefore, it makes no sense to fine-tune the parameters of the model
to try to reproduce particular observations. Instead, the model may be used
to learn about the general behavior of this kind of complex network of 
interactions and mass transfer, and how its behavior depends on the model 
parameters and on the initial and environmental conditions. 
Even if the model has been designed in the simplest possible way,
it was necessary to introduce a relatively large quantity of free parameters
associated to the physical properties of the five gas ``phases''.

Before describing the results of the model it may be useful to advance that 
the behavior of the system 
is chaotic in the sense that its temporal evolution is very sensitive 
to small changes in the initial or external conditions. This chaotic 
behavior is due to the nonlinearity of the included processes and to 
the large number of relationships between the clouds into the tree 
structure. It seems improbable that the inclusion of additional 
processes in the model can result in the avoidance of this chaotic 
behavior. As will be seen, due to this chaotic behavior, the results
point towards a ``non-universal'' Initial Mass Function.  

As a reference, detailed results will be shown for the set of parameters
given in Table~1, and a particular initial and boundary condition.
Due to the chaotic behavior of the system the results for this Reference
Model are not representative of the detailed behavior of the system.
However, this Reference Model is useful to illustrate the general
manner in which the system evolves. After presenting the Reference 
Model results, the effect of variations of model parameters are
summarized. Finally, by changing the initial and external 
condition in the Reference Model, we try to quantify the magnitude and
nature of variations in the IMF. It is to be noticed that in a review
of results concerning the IMF derived from star counts, Scalo (1998)
concludes that there are strong indications of IMF variations, which
do not seem to correlate with obvious environmental conditions like 
metallicity or stellar density. In this sense, Scalo (1998) suggests 
that theoretical IMF models should account for these variations.

\subsection{Results for the Reference Model}

For the set of parameters given in Table~1, 
Fig.~3 shows the evolution
of the total mass in each level (note that there may be as much
as $10^{l-1}$ structures in level $l$). The initial
state corresponds to a totally warm gas cloud of mass
$M_{o}=2\times 10^{5} M_{\sun}$,
a mass accretion $A_{c}=0$, and
an external radiation $\tilde{U}=0.95$.
It can be seen that the warm gas mass
(long dashed line labeled ``{\it wg}'')
decreases as the
condensation of matter to the neutral gas phase progresses
(dotted line labeled ``{\it ng}''). The structures in the low density
molecular level (dotted-long dashed line labeled ``{\it lmg}'')
begin to condense matter when the mass of neutral
gas reaches $\sim 1.5\times 10^{4} M_{\sun}$ at $t\sim 25 Myr$.
This is the time at which the first {\it ng} cloud (i.e. $l=2$ and $j=1$) 
has accumulated enough material to attenuate the external radiation field 
just below the critical value $U_{cri}(3,1)$, allowing the beginning
of the condensation of 
the first {\it lmg} cloud. This time would be reduced if the parameters 
$\phi _2$, $c_o$ (proportional to
the optical depth by unit column density), or $U_3$ are increased.
When the low density molecular gas has reached a large enough mass
($\sim 3\times 10^{4} M_{\sun}$), the high density molecular
gas (dotted line labeled ``{\it hmg}'')
starts to increase its mass and, rapidly,
it begins to condense protostars
(dotted-short dashed line labeled ``{\it ps}'')
at $t\sim 66.3 Myr$. When
the first stars
(solid line labeled ``{\it s}'')
are born, at $t\sim 69 Myr$, both the high density
molecular gas and the protostars are greatly affected, showing
a very irregular behavior associated with
the rapid variation of the radiation density.
The total stellar mass grows up to $\sim 7\times
10^{2} M_{\sun}$ and afterwards remains more or less constant
because the radiation density reaches a ``critical'' value at
which further condensation is inhibited. The decrease of the
radiation density associated to the death of stars is rapidly
compensated by the formation of new stars that again increase
the radiation density. The first star dies at
$t\sim 95 Myr$ and afterwards the mass in remnants
(short dashed line labeled ``{\it r}'') increases continuously.

The corresponding stellar formation history is showed in Fig.~4 (the
dots give the formation events of one or more stars of mass $M$). 
The fact that the model does not produce stars less
massive than $\sim 5 M_{\sun}$ after $t\sim 120 Myr$ may be 
understood in terms of mass accumulation in the intermediate levels.
The radiation of the previously formed stars inhibits the condensation
of matter from the upper levels to the denser ones, accumulating
mass in the low density molecular gas level; but, when 
the radiation field abruptly decreases due to the death of stars,
the protostellar mass rapidly grows allowing the formation of massive
stars only. 
This suppression of formation of low mass stars at late times
is a general property of the model; it remains even when
the model parameters, the initial conditions and the external conditions
are varied (see next section).
It must be mentioned that in this model we are not
considering supernova explosions, that could be an important mechanism
of stimulation or inhibition of star formation due to its drastic effect
on the evolution of the cloud complex.

Figures~5a-5f show how the IMF builds up in a plot of the
number of stars per unit logarithmic mass interval.
At $t=70 Myr$ (Fig.~5a), when the star formation process just began,
the star masses range from $2.5$ to $10 M_{\sun}$; after this first burst,
less massive stars start to form (Figs.~5b-5e, see also Fig.~4) until
$t=120 Myr$ (Fig.~5f), when low mass star formation is inhibited.
At this time,
the IMF shows a maximum at $M_{max}\sim 1.6 M_{\sun}$, and the line
of the better fit for $1.6 \le M \le 10 M_{\sun}$ (showed in the 
Figure~5f like a dashed line) has a slope $\Gamma =-1.48\pm 0.32$.
The efficiency $\epsilon$,
defined as the ratio of the mass transformed in stars to the
initial system mass, is $\sim 0.43 \%$ at $t=120 Myr$. 
But, if at this time,
the efficiency is calculated as the ratio of the mass transformed in 
stars to the gas mass excluding the warm gas, then the efficiency rise
to $\sim 1.2 \%$ (i.e. a factor of about $2.5$ higher). 
The last way for calculating the efficiency of the star formation
efficiency is closer to the way in which observers estimate this quantity.
However, in the rest of the paper, we use the former definition
of $\epsilon$ because it refers to the whole system. 
It must be noticed that for the set of parameters used in the Reference  Model,
the system produces an IMF that has a 
flattening around $1.6 M_{\sun}$ and then a turn over at lower masses,
but observational low-mass IMF studies indicate that the IMFs 
continue to rise below $1 M_{\sun}$, just with a flatter slope (maybe $-0.5$).
Also, the characteristic times for evolution seem too long.
These undesired properties may be improved by adjusting the model
parameters (see below). However, the present model is a very schematic one
to try to compare it with specific observations.
Beside these undesired features, which may be produced by the star
formation condition and because multiple stellar systems,
cloud coalescence, and expanding shells have not been
allowed, the global evolutive pattern does not disagree with the
general idea that we expect for the evolution of an ``unperturbed''
molecular complex.

\subsection{Dependence on model parameters}

In order to learn how the results change when the parameters
of the model are varied, in the ``Reference Model'' various of these 
parameters have been changed one by one.
The parameters that affect more the response of the system are:
$C$ (related to the ability of stellar winds to stop accretion),
$t_l= M_{o,l}/\phi_l$ (related to characteristic time to condense mass
from the level $l-1$ in absence of radiation),
and $c_o$ (the assumed
optical depth of a structure with $M_i=1 \, M_{\sun}$ and
$n_l=1 \, cm^{-3}$).

When $C$ is increased the curve $M_{cri}(t_f)$ in Fig.~2 shifts toward
higher masses. Therefore, protostars may accrete mass during more time and
the mean mass of the formed stars increases. Runs of the model with
$C$ from $1/5$ to $5$ times the reference value ($C=100$) show that
as $C$ is increased the maximum of the IMF function shifts from $\sim 1.0$
to $\sim 2 M_{\sun}$, the slope $\Gamma$, calculated in the range 
$1.6 \le M \le 10 M_{\sun}$, 
increases from $\sim -1.3$ to $\sim -0.6$,
and the efficiency $\epsilon$ 
increases from $\sim 0.3$ to $\sim 3 \%$.
The IMF for $C=500$ is shown in Fig.~6a.
However, a small  variation of $C$ produces in general a large variation of 
these ``integral'' quantities but they remain in the above ranges and with the
described tendencies. These results show that it is difficult to develop 
intuition on this kind of model. One may suspect that an increase of $C$
would reduce the star formation efficiency. Because the mean mass of the 
formed stars increases, the radiation produced by unit mass in stars 
also increases.
Therefore, one may expect a reduction of the critical mass in the 
stellar system in order to maintain the radiation density at the critical
level for condensation. 
However, after examination of the results one learns that this effect is
overcome by others like the increased lifetime of protostars,
i.e., before the more massive protostar in a 
given {\it hmg} cloud starts to shine
as a star, its brothers can grow during a longer time.
Additionally,  massive stars die faster.

When the characteristic time $t_l= M_{o,l}/\phi_l$ 
is increased the process of condensation of 
structures at level $l$ slows. 
For example, a decrease of $\phi_4$ (the parameter that allows to vary 
the velocity of mass exchange between the {\it lmg} and {\it hmg} phases)
decreases the mass suply to the {\it hmg} structures.
Runs for the Reference Model show that little mass is represed in the 
{\it hmg} phase, therefore, a decrease of $\phi_4$ indirectly tends to slow 
the the grow rate of protostars. 
Then, the typical protostar grow curve $M_{ps}(t_f)$ in Fig.~2 tends to 
reduce its slope and consequently, the mean mass of stars decreases.
Runs for $\phi_4$ reduced to $1/2$, $1/5$ and $1/10$ the reference value
($5\times 10^3$) show that the mean mass of stars progressively decrease
while the IMF flattens and its maximum shifts toward low masses.
For the extreme case $\phi_4=5\times 10^2$ the IMF (Fig.~6b) does not
show maximum at all and only low mass stars form ($ <3 M_{\sun}$).

When $c_o$ is increased the optical depth by unit
column density in each level
is increased in the same proportion. Therefore, this parameter affects 
drastically the radiative coupling in the system. In order to analyze
the effect of variations of $c_o$ respect to the reference case it is
useful to adjust the external radiation in such a way to maintain 
$\tilde{U}$ at the reference value $0.95$.
When $c_o$ is set to half the reference value,
the time that the system must wait to form stars is enlarged appreciably
($\sim 200 Myr$) and the IMF is dominated by massive stars. 
Conversely, when $c_o$ is twice the reference value, the time to
start star formation 
is reduced to $\sim 30 Myr$ and low and intermediate mass stars 
dominate the IMF as shown in Fig.~6c. In some 
ways this is the most satisfactory
model with respect to observations. The timescale is reasonable ($30 Myr$)
and the turnover in the IMF is at low masses. 

Even when general trends in the system response can be established 
when a model parameter is varied, it is to be noticed that the
detailed evolution of the system, and even the ``integral quantities'' 
such as the IMF and $\epsilon$, are very sensible to small variations.

\subsection{Dependence on initial and external conditions}

As initial condition for all the runs we have assumed that 
a mass $M_o$ is in the {\it wg} phase and that all the remaining phases are 
empty. This is the most primitive initial condition 
in the sense that it is the configuration that takes more time
to start the formation of stars. This is an unlikely situation because this
homogeneous state hardly occurs in the ISM. However,
this initial condition allows to see how the IMF builds up departing
from a diffuse unstructured gas cloud.
Figure~7 shows the resulting star formation histories
for four values of $M_o$.
The label on each panel indicates the initial mass of the {\it wg} cloud 
in units of $10^5 M_{\sun}$. 
The parameters used are those of the Reference Model except that 
for each $M_o$ the external radiation is adjusted in such a way 
to maintain  $\tilde{U}$ at the reference value $0.95$.
Note that as the initial mass increases the time between bursts decreases,
and the mean mass of formed stars increases.  
Figure~8 shows the corresponding IMFs at $150 Myr$ for the same initial 
masses used in Fig.~7.
The variation of the slope $\Gamma$ of the IMF is evident in Fig.~9, where 
the results for 40 different initial masses are shown.
The plotted values and errors are obtained by fitting a
line in the range $1.6 \le M \le 10 M_{\sun}$ of the IMF
at $150 Myr$.
 
On the other hand, molecular complex are subjected to a variety
of external conditions. Moreover, during its evolution 
the external conditions are expected to vary (i.e. heat input,
external pressure and mass exchange). However, it is assumed
that these conditions remain constant during runs of the model. 
In any case, it is expected that the chaotic behavior of the system
will remain and even increase by adding random variations of 
external conditions. 

Because mass exchange in the system is controlled by radiation, the
external radiation $\tilde{U}$ plays a central role.
As $\tilde{U}$ is decreased the general tendency is that the time 
between bursts decreases and the mean mass of formed stars increases.
It must be noted that in some cases all the mass is 
evaporated to the warm gas phase in the period where only massive stars
are formed. When the radiation field in many places is
high enough to evaporate surrounding high density material,
the opacity decreases and radiation can easily permeate the
tree structure up to the low density levels. In this case, all
the gas becomes warm and $U_{in}(wg)+U_{ext}(wg)>1$.
To restart condensation it is necessary to wait until enough stars
die to decrease the radiation field and the condition
$U_{in}(wg)+U_{ext}(wg)<1$ is again satisfied.

Again, small changes on $\tilde{U}$
produce large changes in the system response as showed in 
Fig.~10 where the evolution of the mass distribution, the 
star formation histories and the corresponding IMFs are plotted
for two close values of $\tilde{U}$.  
The efficiency of the star formation process as function of
$\tilde{U}$ is showed in Figure~11.  
Again, small variations of $\tilde{U}$ produce large variations 
of efficiency of star formation, but as expected, $\epsilon=0$ 
for the threshold value $\tilde{U} = 1$.
The slope $\Gamma$ in the range $1.6 \le M \le 10 M_{\sun}$
of the IMF as function of $\tilde{U}$ is showed in 
Figure~12. The values of $\Gamma$ did not show a correlation
with variations of the external radiation field,
ranging its value from $-0.5$ to
$-2.0$ about of a mean value $\Gamma \simeq -1.4$.

Finally, the statistics of the $\Gamma$ values for $80$ runs
(i.e., $40$ varying the initial mass (Fig.~9) and $40$ varying the
external radiation field (Fig.~12)) is showed in Fig.~13. The 
variability of the model IMFs is evident and roughly agree with
the variability derived in a compilation of observational data by
Scalo (1998).

\section{Conclusions}

A lot of theoretical work has been done
on the IMF (see, for example, Nakano et al. 1995; Adams \& Fatuzzo
1996; Padoan et al. 1997; see also the review of Zinnecker et 
al. 1993).
Many theoretical works have successfully obtained the desired 
power law IMF  departing from 
predefined cloud core properties distributions based on 
numerical and observational results. 
Also, pure statistical or geometrical approaches to model the process of 
gravitational fragmentation have been attempted by several authors
(e.g. Auluck \& Kothari 1954; Reddish 1962; Fowler \& Hoyle 1963;
Larson 1972; Elmegreen \& Mathieu 1983; Di Fazio 1986, Elmegreen 1997).
However, the basic principles that determine
the IMF shape and its dependence on the initial and external
conditions remain unclear.
In a recent review
of results concerning the IMF derived from star counts, Scalo (1998)
concludes that there are strong indications of IMF variations, which
do not seem to correlate with obvious environmental conditions like 
metallicity or stellar density. Even when various sources of 
uncertainties are present in the determination of the IMF, the
results summarized in his Fig. 5 (where the IMF slope for the 
reviewed Milky Way and LMC clusters has been plotted as a function of 
the average value of $\log \, M$) indicate that there is no evidence 
for a clustering of points around the Salpeter value $(-1.3)$.
Instead, there are significant variations of the IMF index at all
masses above $1 \, M_{\sun}$. In this sense, Scalo (1998) suggests 
that theoretical IMF models should account for these variations.

By using a hierarchical structure description, a mass exchange model
has been constructed to simulate some aspects of the evolution of molecular
complexes.
The schematic global approach in this work is completely 
deterministic 
and has allowed us to shift the assumed initial conditions from 
the distribution of the properties in cloud cores to the initial state 
of a single diffuse warm cloud.
Although many of the physical mechanisms that are known to operate in 
molecular complexes are missing or schematized, this kind of models may
help to identify relevant ingredients and typical behaviors, 
and to study the relationship between the various scales involved.
The model presented here can be seen as an example of how this kind
of schematic systems can provide insight about the behavior of
complex ISM structures.  

Even when the model considers few physical processes, its
hierarchical structure allows to follow the evolution of a large 
quantity of variables at various scales. 
In fact, for a 10 children structure tree of five levels,
at each time step the model actualizes
the mass and the radiation density of 11111 substructures,
and the condition for the formation or death of stars in $10^4$
places is verified.
Then, the results showed are  only a tiny fraction of the 
information generated in a run of the model. 

It would be happen that a fine tuning of the parameters would be
necessary to obtain a ``reasonable'' evolution; i.e. a progressive
condensation of diffuse phases into denser ones through
the tree structure -  star formation at few places - evaporation of
dense structures close to these places - star formation at new places
until the global radiation field inhibits star formation at any place -
new episodes of star formation when the radiation field decreases due
to the death of stars.
However, the model is robust in the sense that when the basic 
parameters or the initial and external conditions are varied the 
behavior of the system maintains its general pattern. 
This robustness comes from a global regulation that is controlled by the
radiation field; that is: condensation tends to be inhibited 
by the radiation from
stars but stars form from condensations.
Molecular complexes are common structures that occurs
in a variety of circumstances and therefore the robustness of the model
may be seen as a desirable feature. 
It should be mentioned that the model tends to produce an IMF
that have a maximum at too high mass and also the characteristic times
for evolution seem too long. These undesired properties may be
improved by adjusting the model parameters as shown in Sec.~3.2.
For example, increasing the optical depth parameter $c_o$ resulted
in a reduced time scale for the star formation of only $30 Myr$ and
an IMF that peaked at lower mass than the standard model.
The deficit of high mass stars in that model may reflect the need for 
an accounting of collisional coalescence in the present model.
However, too many processes are obviated in the model
to try to compare its results with observations.
Additionally, it makes no sense to adjust the model parameters without
inferring their values from observations.
The condition for star formation plays an important role in determining
the shape of the IMF, and therefore the
undesirable features in the IMF may also be improved by modifying
the Adams \& Fatuzzo's condition, and by allowing the
possibility of formation of multiple stellar systems, coalescence, 
stellar winds, HII regions, and explosions.

Notwithstanding the simplifications involved, this deterministic model 
reveals an essential feature that likely will remain if
additional physical processes are included.
That is: {\it {
The detailed behavior of the system is very sensitive to the
variations on the initial and external conditions, suggesting that
a ``universal'' IMF is very unlike.}} 
It is interesting to note that looking at the ensemble of results
presented here, the slope $\Gamma$
of the IMF at high masses (Fig.~13) shows 
a distribution
that shares some characteristics with that derived from observational data
(Scalo 1998).
Obviously, this conclusion refers to the IMF for a single star forming
``cloud''. When talking about the universality of the IMF it is necessary
to specify the scale at which the IMF refers. 
When an average over an ensemble of star forming
regions is used to construct the IMF, the differences for different
ensembles tend to be reduced. Then studies that use this kind of average and
point toward a ``universal'' IMF are not
necessarily in contradiction
with a ``non-universal'' IMF at the scale of single star forming 
regions (i.e. Scalo 1998, and this work). 
Although it is out of the scope of the present work, the model can be
used to quantify the variation of the average IMF for an ensemble of runs when
a model parameter is varied.

Finally, the model produces a characteristic pattern of evolution 
that is reminiscent of some aspects expected to occur in real 
molecular complexes.
This fact, together with the robustness of the model,
suggests that the included physical processes 
(phase transitions regulated by the radiation field) may play a role 
in the global evolution of molecular complexes.
 
\acknowledgments
We are very grateful to the referee Prof. John Scalo for the
many suggestions that helped us to put our model in a
much better context. 
This work was supported by C.D.C.H.T. of the
Universidad de Los Andes and
by CONDES of the Universidad del Zulia, Venezuela.

\clearpage

\figcaption{Schematic representation of mass flux in the tree
(see text). For simplicity only two children for each cloud have
been represented.}

\figcaption{The critical mass for star formation ($M_{cri}$)
as a function of the formation time ($t_{f}$). The continuous line
is the assumed functional form (eq.~10),
the dashed line shows the condition obtained from Adams \& Fatuzzo (1996),
and the dotted line shows a typical mass evolution of a
protostar ($M_{ps}$).}

\figcaption{The evolution of the total mass of warm gas ({\it wg}),
neutral gas ({\it ng}),
low density molecular gas ({\it lmg}), high density molecular
gas ({\it hmg}), protostars ({\it ps}), stars ({\it s}) and remnants ({\it r}),
for the Reference Model (see text) for
$M_{o} = 2\times 10^{5} M_{\sun}$ and
$\tilde{U} = 0.95$.}

\figcaption{Star formation events for the parameters
values used in Figure~3.}

\figcaption{The IMF at various times for the same parameter values
used in Fig.~3. a) $t=70 Myr$,
b) $t=80 Myr$, c) $t=90 Myr$, d) $t=100 Myr$, e) $t=110 Myr$
and f) $t=120 Myr$. The dashed line in Fig.~5f shows
the line of the better fit for $1.6 \le M \le 10 M_{\sun}$.}

\figcaption{
The IMF at $t=120 Myr$ when the parameters
of the Reference Model are varied.
a) when $C$ is five times greater than the reference value;
b) when $\phi_4$ is ten times lower than the reference value;
c) when $c_o$ is twice the reference value.
}

\figcaption{
The star formation histories
for four values of the initial mass. The parameters used
are those of the Reference Model and $\tilde{U} = 0.95$.}

\figcaption{
The IMF at $t=150$ for the same initial
masses used in Fig.~7.}

\figcaption{
The slope of the IMF in the range $1.6 \le M \le 10 M_{\sun}$
as function of the initial mass for 40 model runs. The parameters used
are those of the Reference Model and $\tilde{U} = 0.95$.}

\figcaption{
The system response for two close values of $\tilde{U}$
when the parameters of the Reference Model are used, and 
$M_o=2 \times 10^5 M_{\sun}$.
a,b,c) For $\tilde{U} = 0.96$, the evolution of the mass distribution, 
the star formation history, and the IMF at $t=150$, respectively.
d,e,f) The same as in Fig.~10a,b,c but for $\tilde{U} = 0.9625$.}

\figcaption{
The efficiency of the star formation process as function of
$\tilde{U}$ for 40 model runs. The parameters used
are those of the Reference Model and $M_o=2 \times 10^5 M_{\sun}$.}

\figcaption{The slope of the IMF in the range $1.6 \le M \le 10 M_{\sun}$
as function of $\tilde{U}$ for 40 model runs in Fig~11.}

\figcaption{
Histogram of the distribution of the IMF slopes in the range 
$1.6$ to $10 M_{\sun}$ for $80$ runs,
i.e., $40$ varying the initial mass (Fig.~9) and $40$ varying the
external radiation field (Fig.~12).}

\clearpage

\begin{deluxetable}{cccccc}
\footnotesize
\tablecaption{Parameters of the Reference Model}
\tablehead{
\colhead{Level ($l$)} & \colhead{$n_{l}$ ($cm^{-3}$)} &
\colhead{$M_{o,l}$ ($M_{\sun}$)} & \colhead{$\phi_{l}$ ($M_{\sun}/Myr$)} &
\colhead{$U_{l}$} & \colhead{$a_{l}$}
}
\startdata
$1$ &$2\times 10^{-1}$ &\nodata &\nodata &\nodata &\nodata \nl
$2$ &$6.6$ &$1\times 10^{4}$ &$1.5\times 10^{4}$ &$1$ &$8\times 10^{-3}$\nl
$3$ &$2\times 10^{2}$ &$1\times 10^{3}$ &$5.0\times 10^{3}$
&$5\times 10^{-1}$ &$8\times 10^{-3}$\nl
$4$ &$6\times 10^{3}$ &$1\times 10^{2}$ &$5.0\times 10^{3}$
&$2\times 10^{-4}$ &$8\times 10^{-3}$\nl
$5$ &$1\times 10^{5}$ &$1$ &$5.0$ &$1\times 10^{-4}$ &$8\times 10^{-2}$\nl
\enddata
\end{deluxetable}

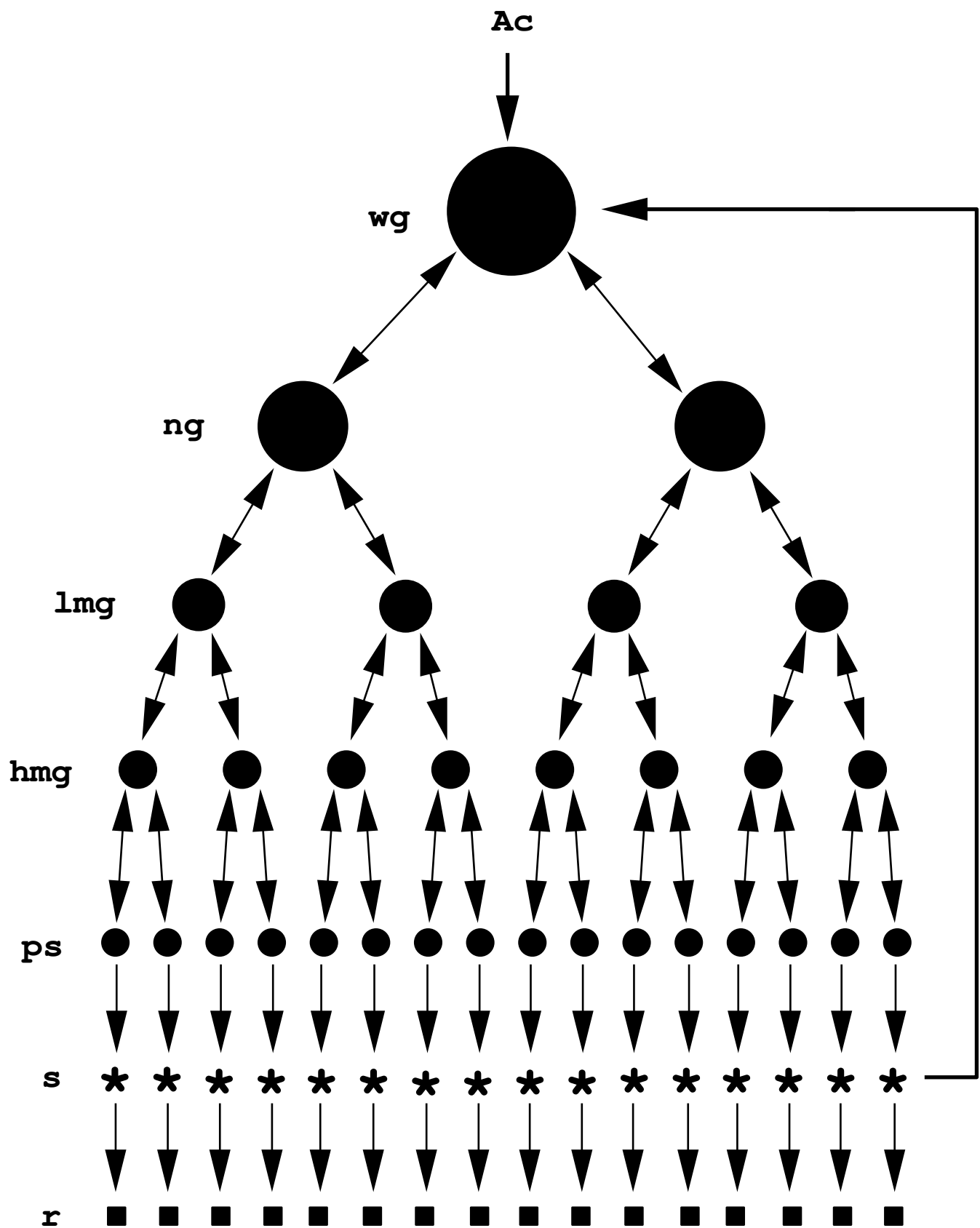
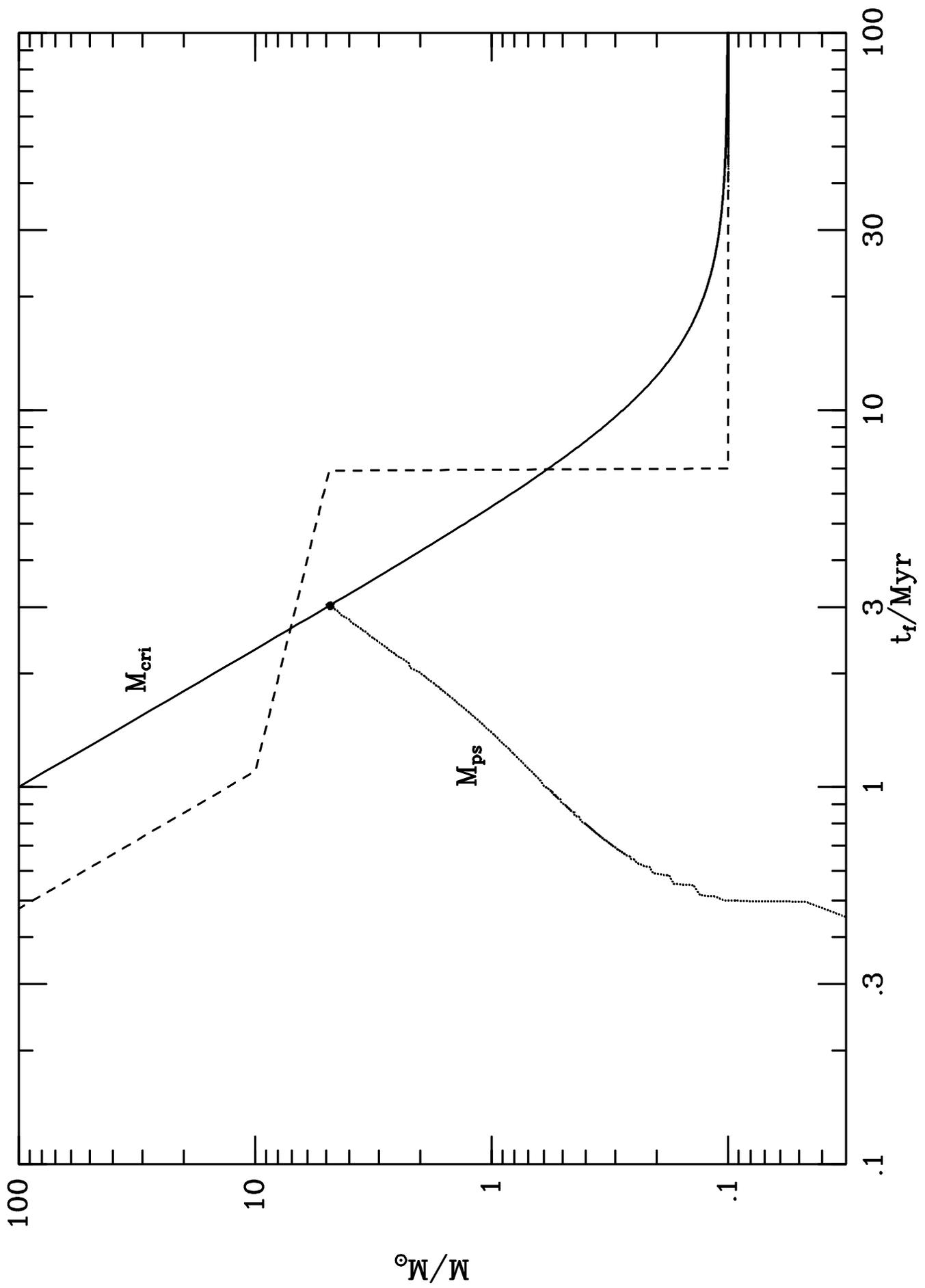
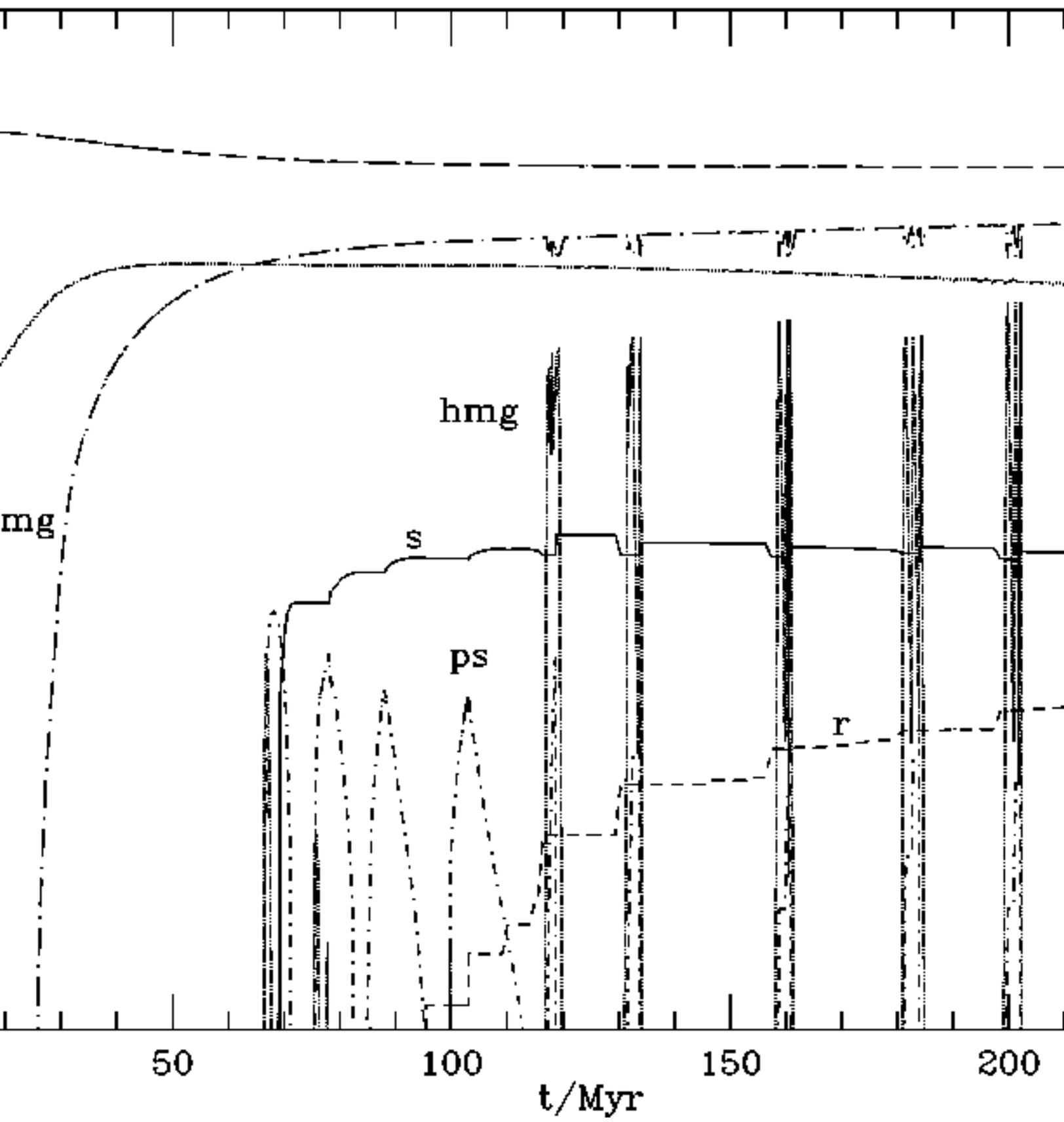
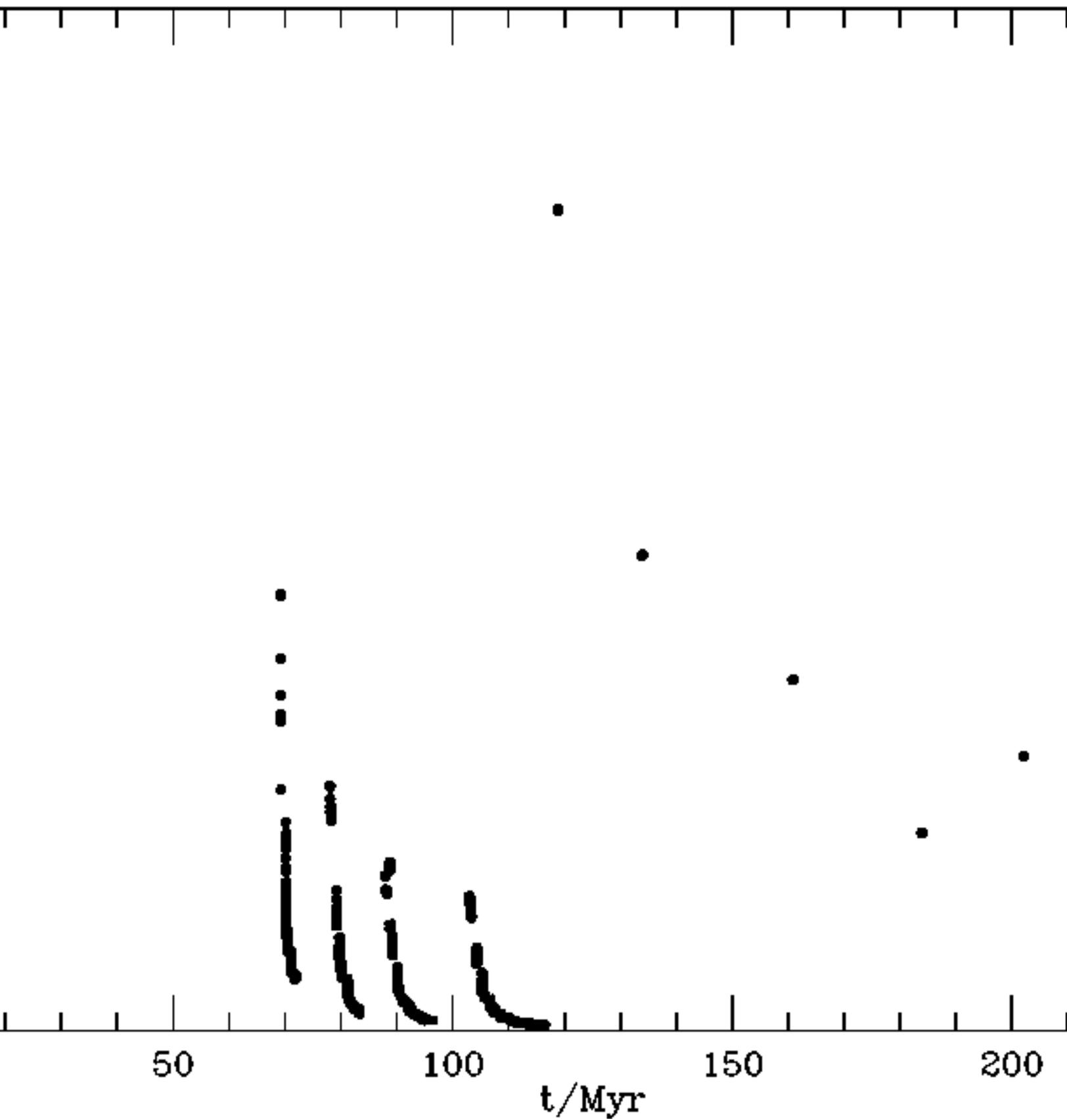
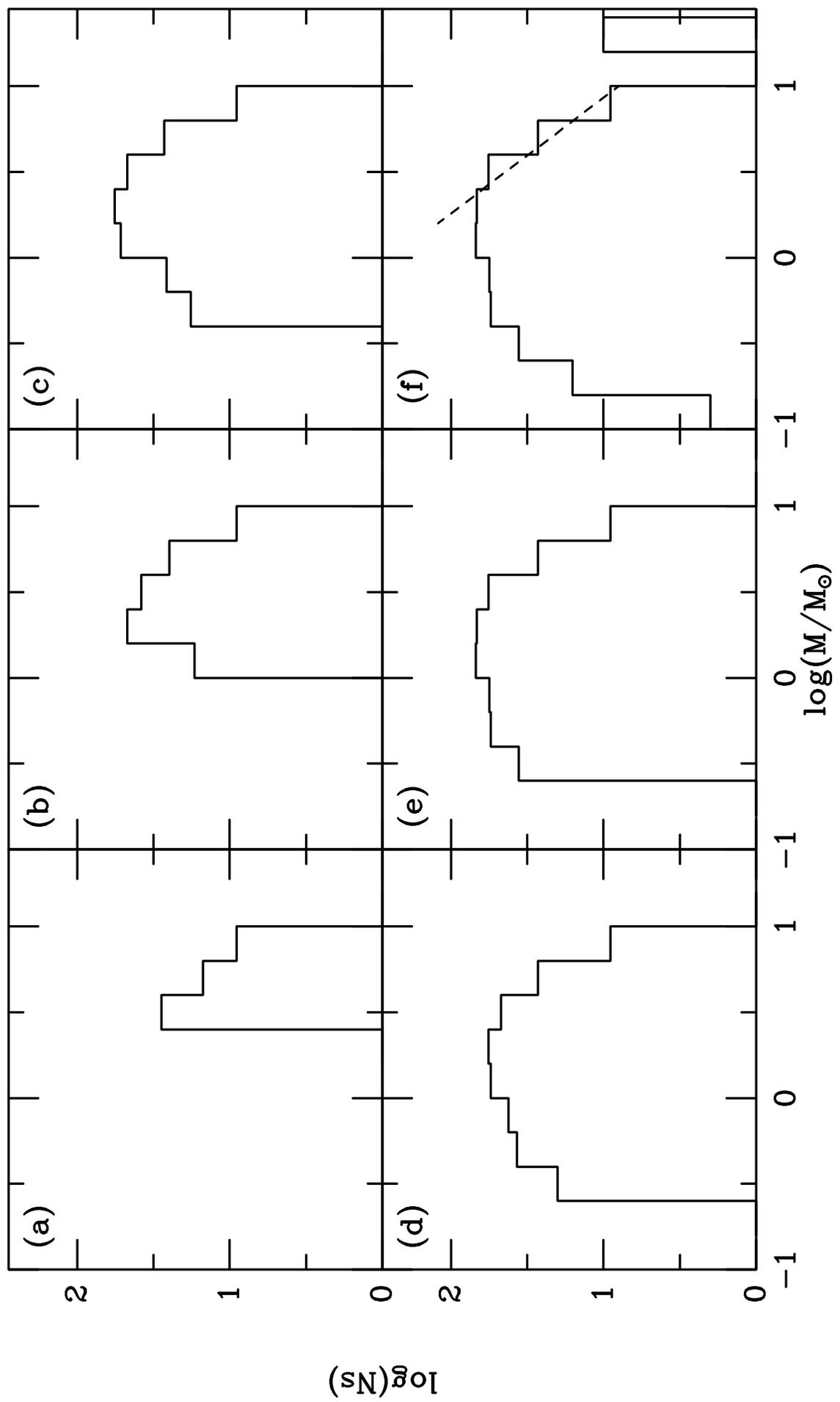
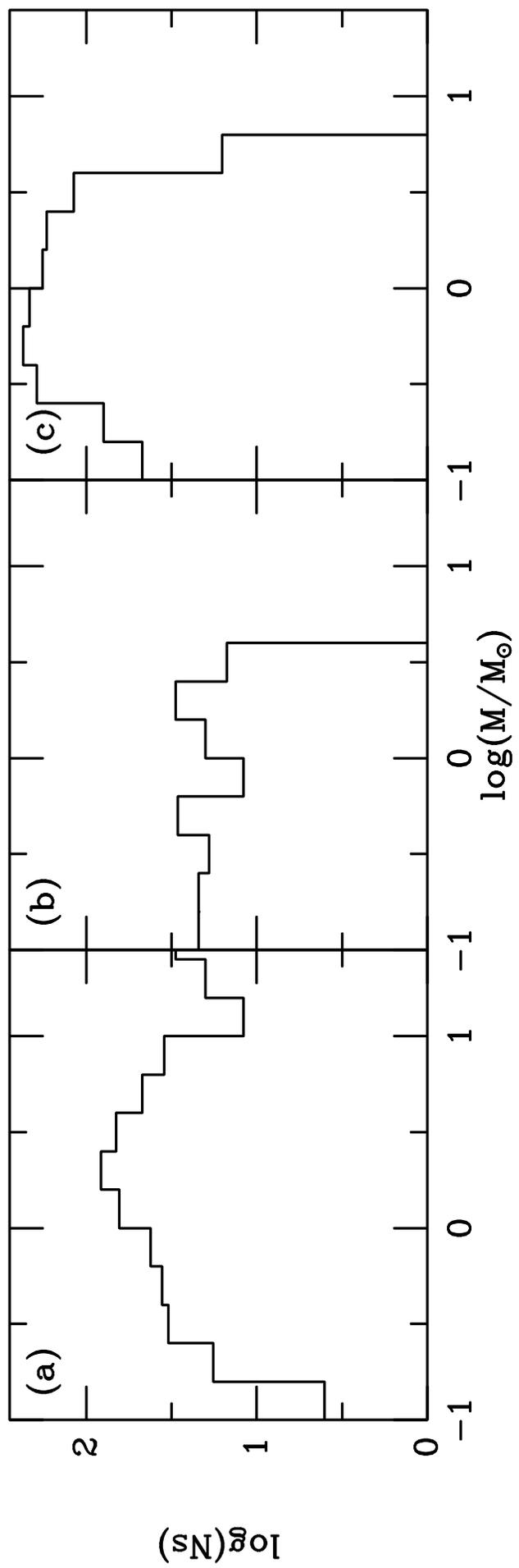
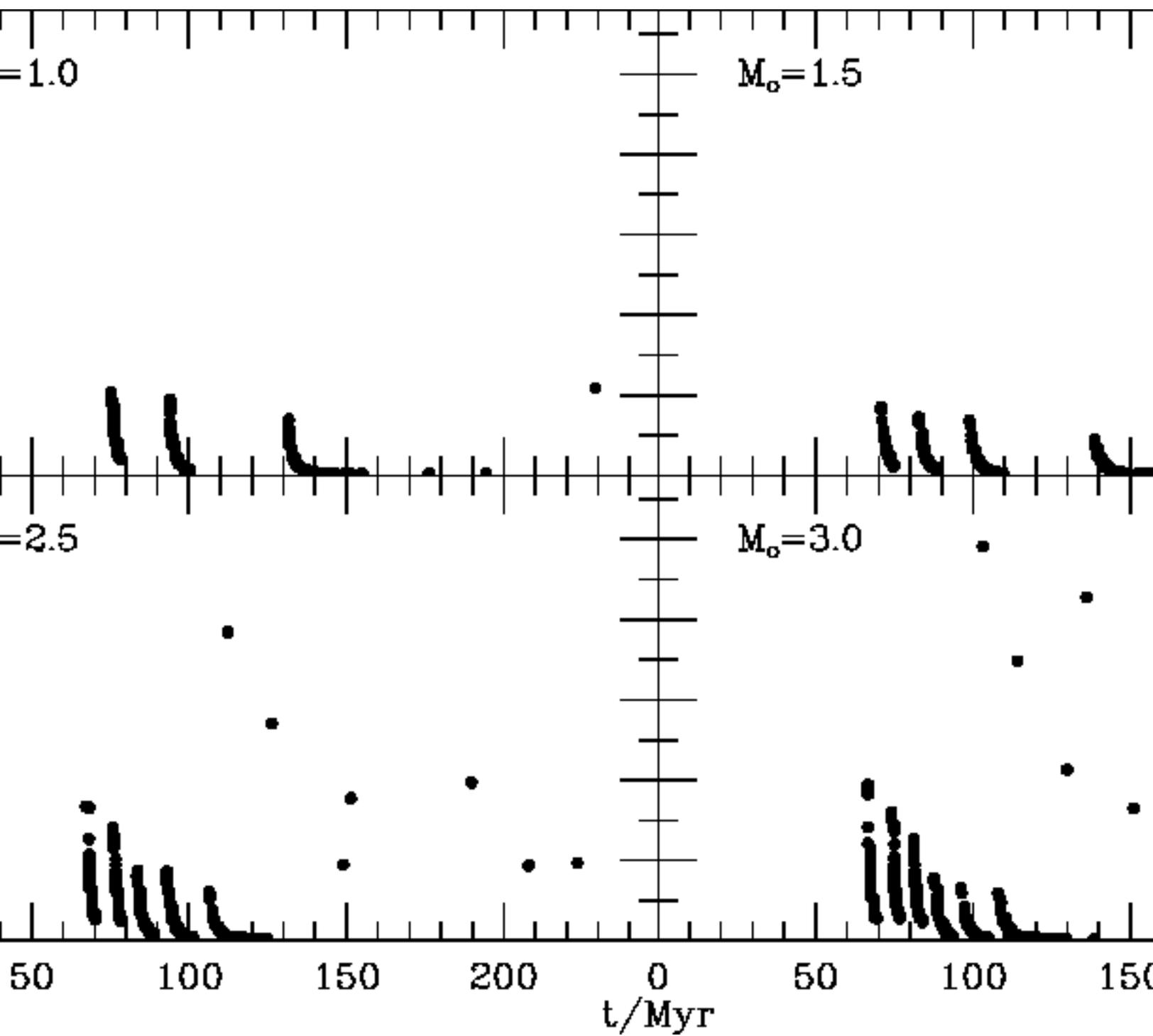
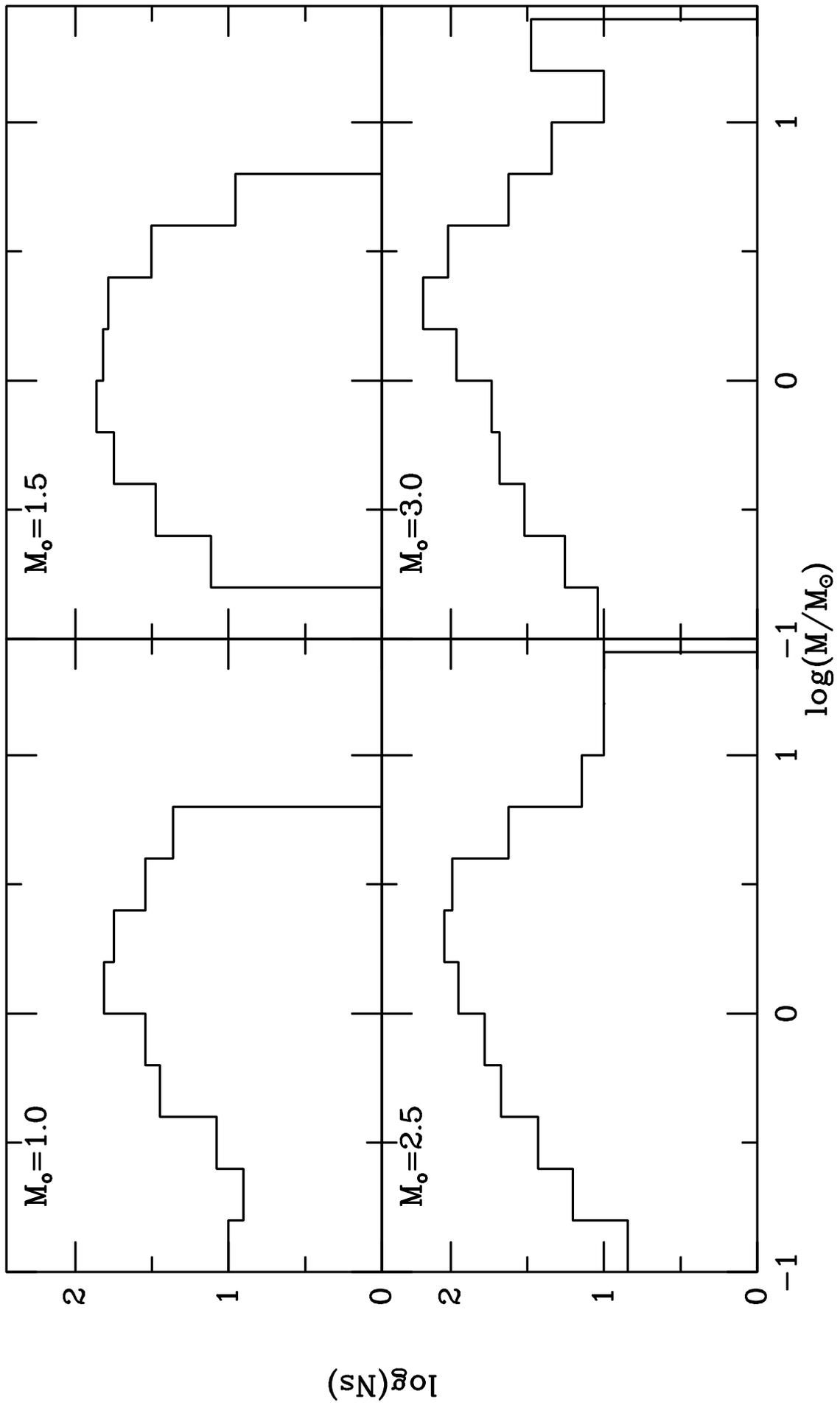
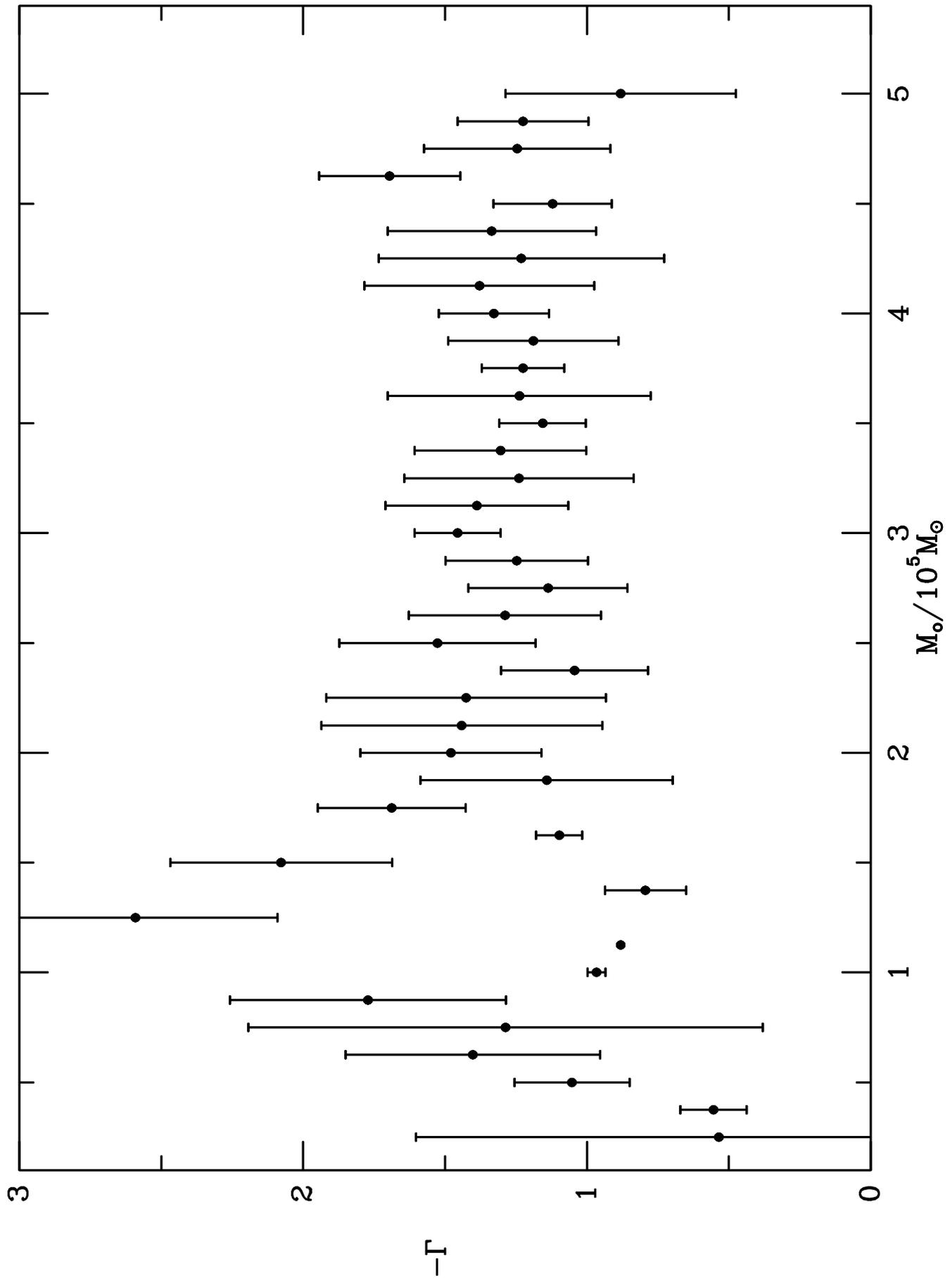
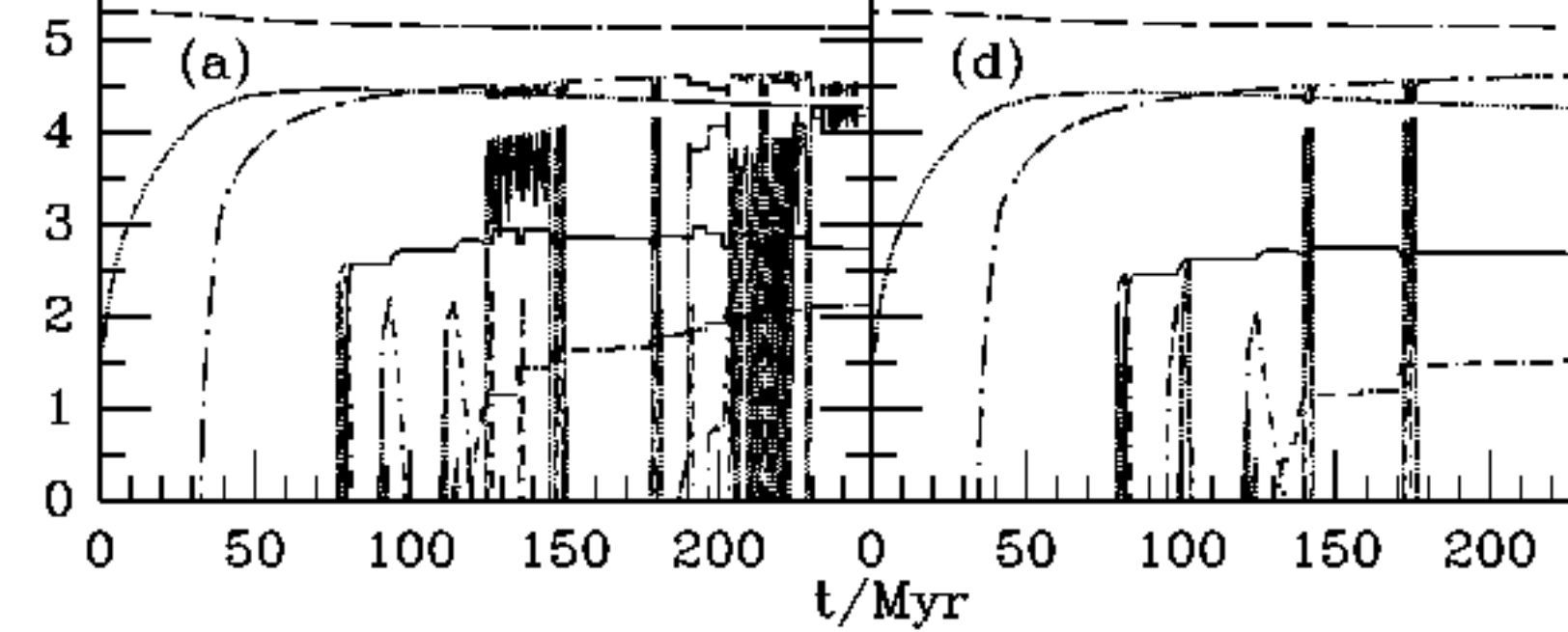
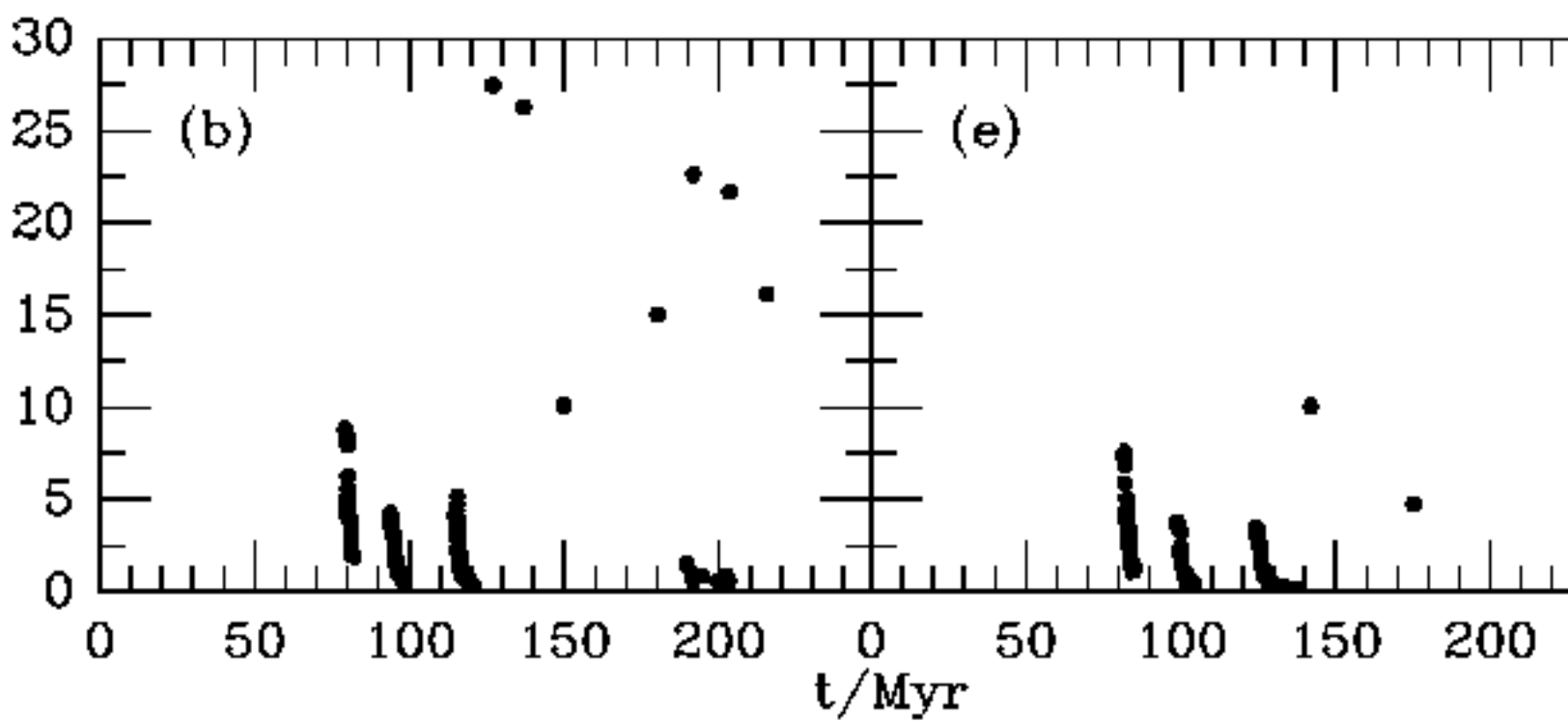
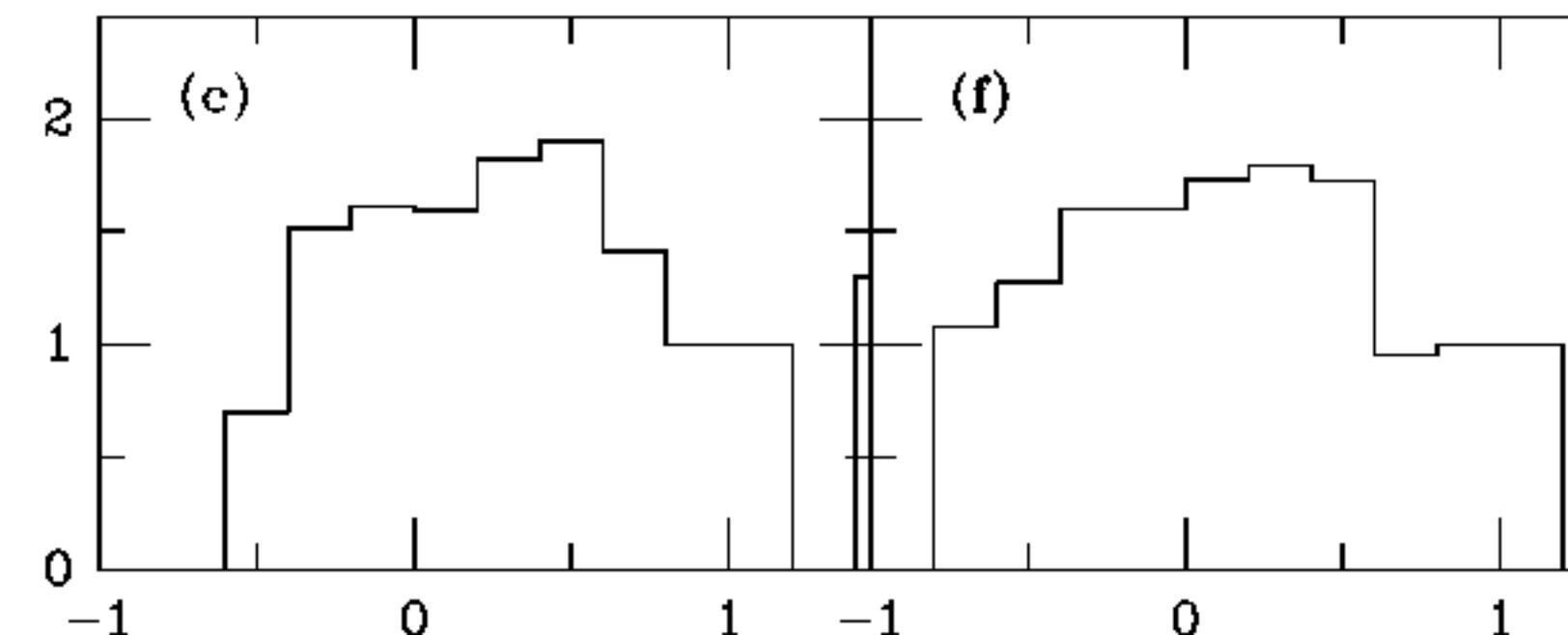
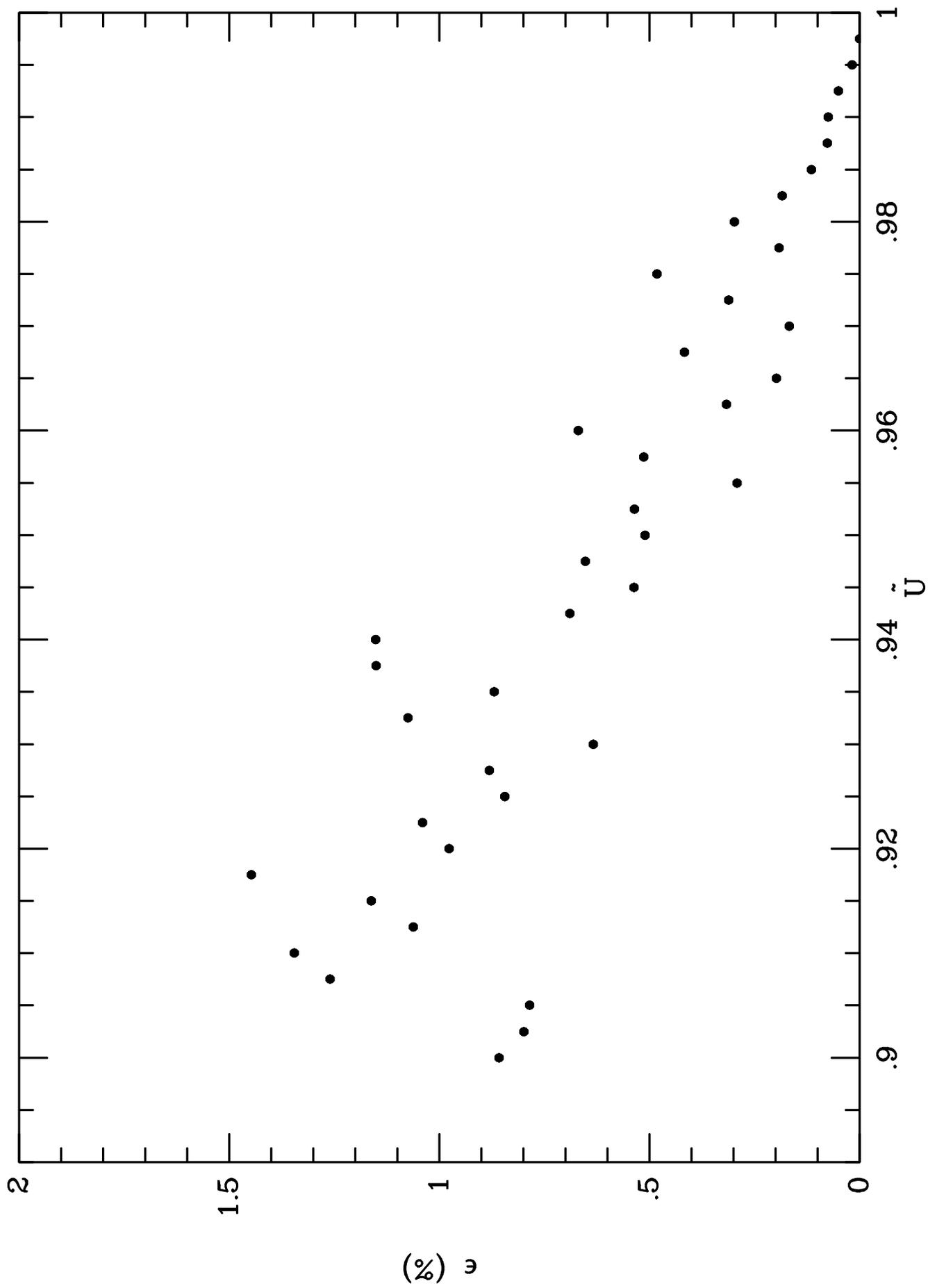
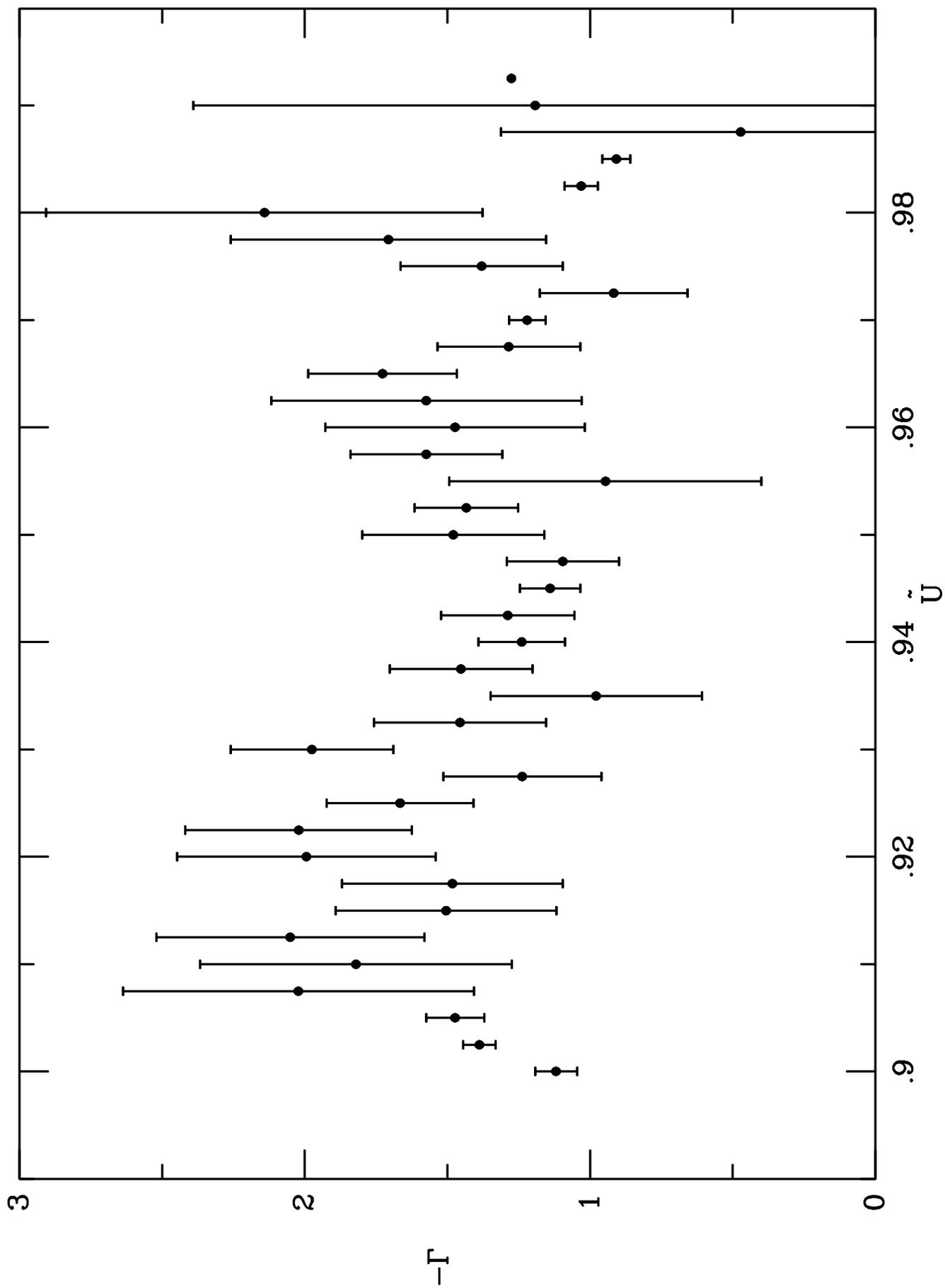
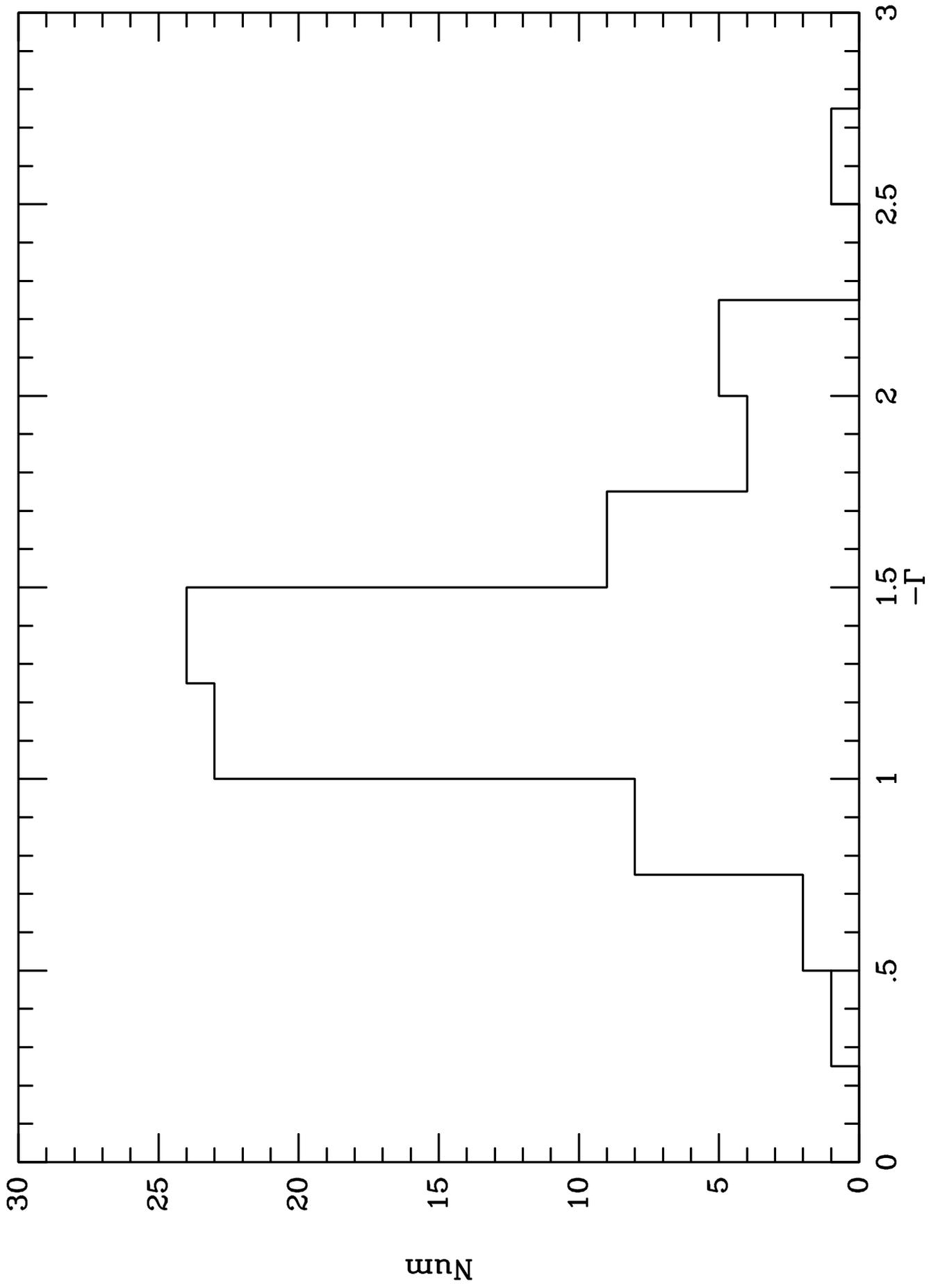
\end{document}